\documentstyle[epsf,12pt,dina4p]{article}

\newcommand{\gsim}{\mbox{\raisebox{-0.4ex}
{$\;\stackrel{>}{\scriptstyle \sim}\;$}}}
\newcommand{\lsim}{\mbox{\raisebox{-0.4ex}
{$\;\stackrel{<}{\scriptstyle \sim}\;$}}}

\begin{document}

\vspace{1 cm}

\begin{titlepage}

\title{
{\bf Measurement of Jet Shapes \\ in Photoproduction at HERA} \\
\author{ZEUS Collaboration } }

\date{ }
\maketitle

\vspace{5 cm}

\begin{abstract}

 The shape of jets produced in quasi-real photon-proton collisions at 
centre-of-mass energies in the range $134-277$~GeV has been measured using the 
hadronic energy flow. The measurement was done with the 
ZEUS detector at HERA. Jets are identified using a cone algorithm in the 
$\eta - \phi$ plane with a cone radius of one unit. Measured jet shapes both 
in inclusive jet and dijet production with transverse energies 
$E^{jet}_T>14$~GeV are presented. The jet shape broadens as the jet 
pseudorapidity ($\eta^{jet}$) increases and narrows as $E^{jet}_T$ increases. 
In dijet photoproduction, the jet shapes have been measured separately for 
samples dominated by resolved and by direct processes. Leading-logarithm 
parton-shower Monte Carlo calculations of resolved and direct processes 
describe well the measured jet shapes except for the inclusive production of 
jets with high $\eta^{jet}$ and low $E^{jet}_T$. The observed broadening of 
the jet shape as $\eta^{jet}$ increases is consistent with the predicted 
increase in the fraction of final state gluon jets. 

\end{abstract}

\vspace{-18cm}
{\noindent
 DESY 97-191 \newline
 October 1997}

\setcounter{page}{0}
\thispagestyle{empty}
\pagenumbering{Roman}
\def\3{\ss}
\parindent0.cm
\parskip 3mm plus 2mm minus 2mm

\newpage

\begin{center}
{\Large  The ZEUS Collaboration}
\end{center}

  J.~Breitweg,
  M.~Derrick,
  D.~Krakauer,
  S.~Magill,
  D.~Mikunas,
  B.~Musgrave,
  J.~Repond,
  R.~Stanek,
  R.L.~Talaga,
  R.~Yoshida,
  H.~Zhang  \\
 {\it Argonne National Laboratory, Argonne, IL, USA}~$^{p}$
% ...
\newpage
% ...
$^{  23}$ now at ATLAS Collaboration, Univ. of Munich \\
% ...
$^{  32}$ partially supported by the Foundation for German-Russian
Collaboration
DFG-RFBR \\ \hspace*{3.5mm} (grant no. 436 RUS 113/248/3 and no. 436 RUS
113/248/2)\\
% ...
%
% \par         % if index listing & table fit to 1 page, put gap here
\newpage   % alternatively: go to newpage, if page is too small
% \institute_references_start    % do not touch or move this line !
%
\begin{tabular}[h]{rp{14cm}}
$^{a}$ &  supported by the Natural Sciences and Engineering Research
          Council of Canada (NSERC)  \\
$^{b}$ &  supported by the FCAR of Qu\'ebec, Canada  \\
$^{c}$ &  supported by the German Federal Ministry for Education and
          Science, Research and Technology (BMBF), under contract
          numbers 057BN19P, 057FR19P, 057HH19P, 057HH29P, 057SI75I \\
$^{d}$ &  supported by the MINERVA Gesellschaft f\"ur Forschung GmbH,
          the German Israeli Foundation, and the U.S.-Israel Binational
          Science Foundation \\
$^{e}$ &  supported by the German Israeli Foundation, and
          by the Israel Science Foundation
  \\
$^{f}$ &  supported by the Italian National Institute for Nuclear Physics
          (INFN) \\
$^{g}$ &  supported by the Japanese Ministry of Education, Science and
          Culture (the Monbusho) and its grants for Scientific Research \\
$^{h}$ &  supported by the Korean Ministry of Education and Korea Science
          and Engineering Foundation  \\
$^{i}$ &  supported by the Netherlands Foundation for Research on
          Matter (FOM) \\
$^{j}$ &  supported by the Polish State Committee for Scientific
          Research, grant No.~115/E-343/SPUB/P03/002/97, 2P03B10512,
          2P03B10612, 2P03B14212, 2P03B10412 \\
$^{k}$ &  supported by the Polish State Committee for Scientific
          Research (grant No. 2P03B08308) and Foundation for
          Polish-German Collaboration  \\
$^{l}$ &  partially supported by the German Federal Ministry for
          Education and Science, Research and Technology (BMBF)  \\
$^{m}$ &  supported by the Fund for Fundamental Research of Russian Ministry
          for Science and Edu\-cation and by the German Federal Ministry for
          Education and Science, Research and Technology (BMBF) \\
$^{n}$ &  supported by the Spanish Ministry of Education
          and Science through funds provided by CICYT \\
$^{o}$ &  supported by the Particle Physics and
          Astronomy Research Council \\
$^{p}$ &  supported by the US Department of Energy \\
$^{q}$ &  supported by the US National Science Foundation \\
\end{tabular}
%
% \institute_references_end     % do not touch or move this line !

\end{titlepage}

\newpage
\parindent 5mm
\parskip 0mm
\pagenumbering{arabic}
\setcounter{page}{1}
\normalsize

\section{\bf Introduction}

 Photoproduction at HERA is studied via $ep$ scattering at low four-momentum 
transfers ($Q^2 \approx 0$, where $Q^2$ is the virtuality of the exchanged 
photon). In photon-proton reactions, two types of QCD processes contribute to 
jet production at leading order (LO) \cite{owens,drees}: either the photon 
interacts directly with a parton in the proton (the direct process) or the 
photon acts as a source of partons which scatter off those in the proton (the 
resolved process). The first year of HERA operation led to the observation of 
hard scattering in $\gamma p$ collisions with evidence for multijet structure 
as well as the presence of the resolved process \cite{h192,zeoct92}. 
Measurements of dijet events allowed the separation of the resolved and direct 
processes \cite{zenov93}. The jet profiles have been measured and found to be
described by leading-logarithm parton-shower calculations, except for resolved
processes in the forward region \cite{zeoct94,zedij95,h196}.

 In this paper, the internal structure of photoproduced jets is studied at the
hadron level. The investigation of the internal structure of jets gives 
insight into the transition between a parton produced in a hard process and 
the experimentally observable spray of hadrons. In the present study, jets are 
searched for with an iterative cone algorithm \cite{cone2,snow} with radius 
$R=1$ in the pseudorapidity\footnote{The ZEUS coordinate system is defined as 
right-handed with the $Z$-axis pointing in the proton beam direction, 
hereafter referred to as forward, and the $X$-axis horizontal, pointing 
towards the centre of HERA. The pseudorapidity is defined as 
$\eta = - \ln(\tan\frac{\theta}{2})$, where 
the polar angle $\theta$ is taken with respect to the proton beam direction.} 
($\eta$) - azimuth ($\phi$) plane. The jet shape is defined as the average 
fraction of the jet's transverse energy ($E^{jet}_T$) that lies inside an 
inner cone of radius $r$ concentric with the jet defining cone \cite{sdellis}:
\begin{equation}
  \psi(r) = \frac{1}{N_{jets}} \sum_{jets} \frac{E_T(r)}{E_T(r=R)}
\end{equation}
where $E_T(r)$ is the transverse energy within the inner cone of radius $r$
and $N_{jets}$ is the total number of jets in the sample. By definition,
$\psi(r=R)=1$. 

 The jet shape is affected by fragmentation and gluon radiation. However,
at sufficiently high $E^{jet}_T$ the most important contribution is predicted
to come from gluon emission off the primary parton and, therefore, is
calculable in perturbative QCD. The lowest-non-trivial order contribution 
to the jet shape is given by next-to-leading order (NLO) QCD calculations for 
the reaction $AB \rightarrow {\rm jet}+X$. 
Perturbative QCD predicts that gluon jets are broader than quark jets as a 
consequence of the fact that the gluon-gluon is larger than the quark-gluon 
coupling strength.

 Measurements of jet shapes have been made in $\bar{p}p$ collisions at 
$\sqrt{s}=1.8$~TeV using only charged particles \cite{cdf1} as well as both 
neutral and charged particles \cite{d01}, and a qualitative agreement with 
NLO QCD calculations \cite{sdellis,jetrad} was found. Similar measurements 
have been made in $e^+e^-$ interactions at LEP1 using both neutral and charged 
particles \cite{opal1} and found to be well described by leading-logarithm 
parton-shower Monte Carlo calculations. It was observed \cite{opal1} that the 
jets in $e^+e^-$ are significantly narrower than those in $\bar{p}p$. This is
due to the different mixtures of quark and gluon jets in these two 
environments \cite{opal1}. Measurements of the jet width at LEP1 have shown 
that gluon jets are indeed broader than quark jets \cite{lepjewi}.

 In this paper, measurements are presented of the jet shapes in both inclusive 
jet and dijet photoproduction at centre-of-mass energies in the range 
$134-277$~GeV. The data sample used in this analysis was collected with the 
ZEUS detector in $e^+p$ interactions at the HERA collider and corresponds to 
an integrated luminosity of 2.65~pb$^{-1}$. Jets are selected with 
$E^{jet}_T>14$~GeV and $-1<\eta^{jet}<2$. The jet shape is measured using the 
ZEUS calorimeter and corrected to the hadron level.
The measurements are presented as a
function of the jet transverse energy and pseudorapidity. In dijet 
photoproduction, the jet shapes are measured for resolved and direct 
processes defined by the fraction of the photon's momentum 
participating in the production of the two jets of highest 
$E^{jet}_T$. 

 Measurements of jet shapes are compared to QCD calculations based on 
different approaches:
\begin{itemize}
 \item[1)] LO QCD calculation including initial and final state QCD radiation 
       in the leading-logarithm parton-shower approximation as implemented in 
       the program PYTHIA~5.7 \cite{pythia}. The final state parton system 
       is hadronised using the LUND string model \cite{lund}.
 \item[2)] LO QCD calculation as implemented in the program PYTHIA without
       initial and final parton radiation. The final state parton system 
       is hadronised using the LUND string model.
 \item[3)] Fixed-order perturbative QCD calculation of the reaction 
       $e^+ p \rightarrow {\rm 3 \; partons} +  X$~\cite{kramer,kramerf}.
       Fragmentation effects are not included.
\end{itemize}
These comparisons allow the study of the relative importance of parton 
radiation and fragmentation in the formation of a jet as well as the 
differences between quark and gluon jets. It should be noted that the first 
two predictions refer to the hadron level while the third 
refers to the parton level. For the first two predictions, jets are 
searched for in the final state hadronic system using the same jet algorithm 
as in the data. For the third prediction, the experimental jet 
algorithm has been simulated by the introduction of an additional parameter, 
$R_{SEP}$, as proposed in \cite{sdellis}.

\section{Experimental conditions}

 During 1994 HERA operated with 153 colliding bunches of protons of energy 
$E_p=820$~GeV and positrons of energy $E_e=27.5$~GeV, with 96~ns between 
bunch crossings. 

 A description of the ZEUS detector can be found in \cite{sigtot,status}.
The components used in this analysis are briefly discussed. The 
uranium-scintillator calorimeter (CAL) \cite{calori} covers 99.7\% of the 
total solid angle. It consists of the forward calorimeter (FCAL) covering the 
range $2.6^\circ<\theta<36.7^\circ$ in polar angle, the barrel calorimeter 
(BCAL) covering $36.7^\circ~<~\theta~<~129.1^\circ$, and the rear calorimeter 
(RCAL) covering $129.1^\circ<\theta<176.2^\circ$. Holes of $20\times 20$~cm$^2$
in the centre of FCAL and RCAL are required to accommodate the HERA beampipe.
For normal incidence, the depth of the CAL is 7 interaction lengths in FCAL, 
5 in BCAL and 4 in RCAL. Each of the calorimeter parts is subdivided into
towers which in turn are segmented longitudinally into one electromagnetic 
(EMC) and one (RCAL) or two (FCAL, BCAL) hadronic (HAC) sections. The sections 
are further subdivided into cells of inner-face sizes of approximately 
$5 \times 20$~cm$^2$ ($10 \times 20$~cm$^2$ in the RCAL) for the EMC and 
$20 \times 20$~cm$^2$ for the HAC sections. Each cell is viewed by two 
photomultipliers. At $\theta = 90^{\circ}$ the size of an EMC (HAC) cell in 
the $\eta - \phi$ plane is approximately $0.04 \times 11^{\circ}$ 
($0.16 \times 11^{\circ}$). Under test beam conditions the calorimeter has an 
energy resolution of $\sigma/E$~=~18\%/$\sqrt{E ({\rm GeV})}$ for electrons 
and $\sigma/E$~=~35\%/$\sqrt {E({\rm GeV})}$ for hadrons. In order to minimise 
the effects of noise due to the uranium radioactivity, all EMC (HAC) cells 
with an energy deposit of less than 60~(110)~MeV are discarded in the 
analysis. For isolated energy deposits, consisting of one cell surrounded by 
empty cells, this cut was increased to 100~(150)~MeV.
Particles impinging on the CAL lose energy in the inactive material
in front of the CAL. In the region relevant for the present analysis, the 
inactive material constitutes about one radiation length except in the region
around the rear beampipe, $\theta \gsim 170^{\circ}$, and the solenoid support
structure, $25^{\circ} \lsim \theta \lsim 45^{\circ}$ and $130^{\circ} \lsim 
\theta \lsim 145^{\circ}$, where it is up to $2.5$ radiation 
lengths. For the following measurements, the transverse energy and the 
shape of the jets have been corrected for these energy losses (see Section~5).

 The tracking system consists of a vertex detector (VXD) \cite{vxd},
a central tracking chamber (CTD) \cite{ctd}, and a rear tracking 
detector (RTD) \cite{status} enclosed in a 1.43 T solenoidal magnetic 
field. The interaction vertex is measured with a typical resolution along 
(transverse to) the beam direction of 0.4~(0.1)~cm. 

 Proton-gas events occurring upstream of the nominal interaction point
are out of time with respect to the $e^+p$ interactions and are rejected
by timing measurements using a set of scintillation counters.

\subsection{Trigger conditions}

 The ZEUS detector uses a three-level trigger system \cite{status}. At the 
first level events were triggered on a coincidence of a regional or transverse
energy sum in the CAL and at least one track from the interaction point 
measured in the CTD. At the second level a total transverse energy of at least 
$8$~GeV, excluding the energy in the eight CAL towers immediately surrounding 
the forward beampipe, was required, and cuts on CAL energies and timing were 
used to suppress events caused by interactions between the proton beam and 
residual gas in the beampipe.

 The full event information was available at the third-level trigger (TLT). 
Tighter timing cuts as well as algorithms to remove beam-halo muons and cosmic 
muons were applied. For this analysis, the following additional conditions 
were required: a) the event has a vertex reconstructed by the tracking 
chambers with the $Z$ value in the range $|Z|<60$~cm; b) $E-p_Z \geq 8$~GeV, 
where $E$ is the total energy as measured by the CAL, $E=\sum_i E_i$, and 
$p_Z$ is the $Z$-component of the vector $\vec{p} = \sum_i E_i \vec r_i$; in 
both cases the sum runs over all CAL cells, $E_i$ is the energy of the 
calorimeter cell $i$ and $\vec r_i$ is a unit vector along the line joining 
the reconstructed vertex and the geometric centre of the cell $i$; c) 
$p_Z/E \leq 0.95$ to reject beam-gas interactions (this cut was not applied 
for events with $E-p_Z \geq 12$~GeV); and d) the total transverse energy as 
measured by the CAL, excluding the cells with polar angles below $10^{\circ}$, 
exceeds $20$~GeV.

 For studies of the trigger efficiency an additional sample of events was 
selected by the TLT based upon jets found using a cone algorithm with radius 
$R=1$ applied to the CAL cell energies and positions. The events were 
required to fulfill the same conditions a), b) and c) as above, and to have at 
least one jet of transverse energy, as measured by the CAL, 
$E^{jet}_{T,cal}>6.5$~GeV and $\eta^{jet}_{cal} < 2.5$.

\section{\bf Data selection}

 Events from quasi-real photon proton collisions containing jets were selected 
offline using similar criteria as reported previously \cite{zeoct94}. The 
main steps are briefly 
discussed here. A search for jet structure using the CAL cells (see Section~5) 
is performed, and events with at least one jet of transverse energy, as
measured by the CAL, $E^{jet}_{T,cal}>10$~GeV and $-1<\eta^{jet}_{cal}<2$ are 
retained. The contamination from beam-gas interactions, cosmic showers and 
halo muons is negligible after demanding: a) at least two tracks pointing to 
the vertex; b) the vertex position along the beam axis to lie in the range
$-$29~cm~$ < Z < $~36~cm; c) less than five tracks not associated with
the vertex and compatible with an interaction upstream in the direction
of the proton beam; d) the number of tracks not associated with the vertex
be less than 10\% of the total number of tracks; and e) the total missing 
transverse momentum (${p_T\hspace{-3.5mm}\slash\hspace{1.5mm}}$) be small 
compared to the total transverse energy ($E^{tot}_T$) by requiring
${p_T\hspace{-3.5mm}\slash\hspace{1.5mm}} /\sqrt{E^{tot}_T} <
2$~GeV$^{\frac{1}{2}}$. Deep inelastic $e^+p$ scattering (DIS) neutral current
events with an identified scattered positron candidate in the CAL according
to the algorithm described in \cite{zenov93} are removed from the sample.

The selected sample consists of events from $e^+p$ interactions with 
$Q^2 \leq 4$~GeV$^2$ and a median of $Q^2 \approx 10^{-3}$~GeV$^2$. The 
$\gamma p$ centre-of-mass energy ($W$) is calculated using the expression 
$W=\sqrt{y s}$, where $y$ is the inelasticity variable and $s$ is the squared 
$e^+p$ centre-of-mass energy ($300^2$~GeV$^2$). The event sample is
restricted to the kinematic range $0.2 < y < 0.85$ using the following
procedure. The method of Jacquet-Blondel \cite{jacblo}, 
$y_{JB} = (E-p_Z)/(2 E_e)$,
is used to estimate $y$ from the energies measured in the CAL cells. Due to
the energy lost in the inactive material in front of the CAL and to particles
lost in the rear beampipe, $y_{JB}$ systematically underestimates the true $y$
by approximately 20\%, an effect which is adequately reproduced in the Monte 
Carlo simulation of the detector. To compensate for this deficiency, the event 
selection required $0.16 < y_{JB} < 0.7$. The data sample consists of 15,368 
events with a total of 18,897 jets. The only significant remaining background 
is from unidentified DIS neutral current interactions with $Q^2 >$~4~GeV$^2$, 
which is estimated by Monte Carlo techniques to be below 2\%.

\section{\bf Monte Carlo simulation }

 The response of the detector to jets and the correction factors for the jet 
shapes were determined from samples of Monte Carlo events.

 The programs PYTHIA~5.7 \cite{pythia} and HERWIG~5.8 \cite{herwig} were
used to generate photoproduction events for resolved and direct
processes. In PYTHIA the positron-photon vertex was modelled according to the
Weizs\"{a}cker-Williams approximation. In the case of HERWIG,
the exact matrix elements were used for direct processes
($e^+g \rightarrow e^+ q \bar{q}$ and $e^+q \rightarrow e^+ q g$) and the
equivalent photon approximation for resolved processes.
Events were generated using GRV-HO \cite{grv} for the photon parton 
distributions and MRSA \cite{mrsa} for the proton parton distributions. 
In both generators, the partonic processes were simulated using LO matrix
elements, with the inclusion of initial and final state parton showers.
Fragmentation into hadrons was performed using the LUND
string model \cite{lund} as implemented in JETSET \cite{jetset} in the case of
PYTHIA, and the cluster model in the case of HERWIG. Samples of events
were generated with different values of the cutoff on the transverse
momentum of the two outgoing partons, starting at $\hat p_{Tmin}= 8$~GeV.

 Additional samples of events were generated using the option of multiparton
interactions (MI) in PYTHIA. This option, which applies only to resolved 
processes, adds interactions between the partons in the proton and the 
photon remnants calculated as LO QCD processes to 
the hardest scattering process of the event. The PYTHIA MI events were 
generated with a cutoff for the effective minimum transverse momentum for 
multiparton interactions of $1.0$~GeV and with a cutoff on the transverse 
momentum of the two outgoing partons from the hardest scattering of 
$\hat p_{Tmin}= 8$~GeV.

 All generated events were passed through the ZEUS detector and trigger
simulation programs \cite{status}. They were reconstructed and analysed by
the same program chain as the data.

\section{Jet search and reconstruction of the jet shape}

 An iterative cone algorithm in the $\eta-\phi$ plane \cite{cone2,snow} 
(PUCELL) is used to reconstruct jets, from the energy measured in the CAL 
cells for both data and simulated events, and also from the final state
hadrons for simulated events.

  The procedure is explained in detail for the jet reconstruction from the CAL 
cell energies ($cal$ jets). In a first step, each CAL cell with a 
transverse energy in excess of 300~MeV is considered as a seed for the search.
Their corresponding $\eta-\phi$ values are obtained from the unit vectors 
joining the vertex of the interaction and the geometric centres of the cells. 
The seeds are then combined into preclusters starting from that with highest 
transverse energy if their distance in the $\eta-\phi$ plane, 
$\sqrt{(\Delta\eta)^2+(\Delta\phi)^2}$, is smaller than 1 unit. The
axis of the precluster is defined according to the Snowmass convention
\cite{snow}, where $\eta^{precluster}$ ($\phi^{precluster}$) is the 
transverse-energy weighted mean pseudorapidity (azimuth) of all the seeds
belonging to that precluster. 

 In a second step, a cone of radius $R=1$ is drawn around each precluster 
and all the CAL cells within that cone are combined to form a cluster. The
axis of the cluster is defined according to the same prescription as for
the preclusters but including all the CAL cells belonging to that cluster. A 
new cone of $R=1$ is then drawn around the axis of the resulting cluster. All 
cells with geometric centres inside the cone are used to recalculate a new 
cluster axis. The procedure is iterated until the content of the cluster 
remains unchanged.

 In a third step, the energy sharing of overlapping clusters is considered.
Two clusters are merged if the overlapping energy exceeds 75\% of
the total energy of the cluster with the lower energy; otherwise two 
different clusters are formed and the common cells are assigned to the 
nearest cluster. Finally, a cluster is called a jet if the transverse energy
as measured by the CAL, $E^{jet}_{T,cal}$, exceeds 10~GeV. The angular
variables associated with the $cal$ jets are denoted by $\eta_{cal}^{jet}$ and 
$\phi_{cal}^{jet}$.

 The following procedure was used to reconstruct the jet shape from the
CAL cells: for each jet the sum of the transverse 
energies of the CAL cells assigned to the jet with a distance
$r'=\sqrt{(\Delta \eta)^2+(\Delta \phi)^2}$ to the jet axis smaller
than $r$ is determined, $E_{T,cal}(r)$, and divided by $E_{T,cal}(r=1)$. The 
jet shape as measured with the CAL, $\psi_{cal}(r)$, is thus defined as:
\begin{equation}
 \psi_{cal}(r) = \frac{1}{N_{jets}} 
        \sum_{jets} \frac{E_{T,cal}(r)}{E_{T,cal}(r=1)},
\end{equation}
where the sum runs over all the jets in the selected sample and $N_{jets}$
is the total number of jets in the sample. 

 For the Monte Carlo events, the same jet algorithm is also applied to the 
final state particles. In this search, all particles with lifetimes longer 
than $10^{-13}$~s and with polar angles between $5^{\circ}$ and $175^{\circ}$ 
are considered. The jets found are called $hadron$ jets and the variables 
associated with them are denoted by $E^{jet}_{T,had}$, $\eta^{jet}_{had}$, and 
$\phi^{jet}_{had}$. $Hadron$ jets with $E^{jet}_{T,had}>14$ GeV and 
$-1<\eta^{jet}_{had}<2$ are selected. The same jet shape definition as used 
above for the CAL cells is applied to the final state particles in the case of 
simulated events and the resulting jet shape is denoted by 
$\psi^{MC}_{had}(r)$.

 The comparison of the reconstructed jet variables between the $hadron$ and 
the $cal$ jets in simulated events \cite{zeoct94} shows no significant 
systematic shift in the angular variables $\eta^{jet}_{cal}$ and 
$\phi^{jet}_{cal}$ with respect to $\eta^{jet}_{had}$ and $\phi^{jet}_{had}$. 
However, the transverse energy of the $cal$ jet underestimates that 
of the $hadron$ jet by an average amount of $\approx$~16\% with an r.m.s. of 
11\%. The transverse energy corrections to $cal$ jets averaged over the 
azimuthal angle were determined using the Monte Carlo events \cite{zeoct94}. 
These corrections are constructed as multiplicative factors, 
$C(E^{jet}_{T,cal},\eta^{jet}_{cal})$, which, when applied to the $E_T$ of the 
$cal$ jets give the `corrected' transverse energies of the jets, $E^{jet}_{T}=
C(E^{jet}_{T,cal},\eta^{jet}_{cal}) \times E^{jet}_{T,cal}$ \cite{zeoct94}. 
These corrections mainly take into account the energy losses due to the 
inactive material in front of the CAL.

\subsection{Jet shape corrections}

 The jet shapes as measured with the CAL are corrected to the hadron level 
using the Monte Carlo event samples. The corrected jet shapes are denoted by 
$\psi(r)$ and refer to jets at the hadron level with a cone radius of one unit
in the $\eta-\phi$ plane. The measurements are given for jets of corrected 
transverse energy $E^{jet}_T > 14$~GeV and $-1<\eta^{jet} < 2$, and in the 
kinematic region defined by $Q^2 \leq 4$~GeV$^2$ and $134<W<277$~GeV.

 The reconstructed jet shapes are corrected for acceptance and smearing 
effects using the samples of Monte Carlo events of resolved and 
direct processes. The correction factors also take into account the 
efficiency of the trigger, the selection criteria, the purity and efficiency 
of the jet reconstruction, and the effects of energy losses due to the 
inactive material in front of the CAL. The corrected jet shapes are determined 
bin-by-bin as $\psi(r) = F^{MC}_{cal}(r) \cdot \psi_{cal}(r)$, where the 
correction factors $F^{MC}_{cal}(r)$ are defined as 
$F^{MC}_{cal}(r) = \psi^{MC}_{had}(r)/\psi^{MC}_{cal}(r)$. 
$F^{MC}_{cal}(r)$ is determined separately for each interval in 
$\eta^{jet}$ and $E^{jet}_T$. 
 
 For this approach to be valid, the uncorrected jet shapes in the data must 
be described by the Monte Carlo simulations at the detector level. As 
shown later, this condition is in general satisfied by both the PYTHIA and 
HERWIG simulations although some disagreement is observed in the forward 
region in the case of inclusive 
jet production. The latter can be reduced by adjusting the relative 
contribution of direct processes and the fraction of gluon jets in both 
resolved and direct processes. In the simulated events a $cal$ jet is 
classified as a quark or gluon jet depending on the type (quark or gluon) of 
the closest parton from the two-to-two hard subprocess.

 The following procedure was adopted to obtain the best description of the 
uncorrected jet shapes: a) the relative contributions of resolved and direct 
processes were tuned by a fit to the measured RCAL energy distribution in the 
data; a distinct distribution of the energy deposit in the rear direction 
for resolved (direct) processes is expected due to the presence (absence) 
of the photon remnant; b) the fraction of gluon jets in direct and resolved 
processes was adjusted by a fit to the uncorrected jet shape.
This procedure was applied separately for each $\eta^{jet}$ 
and $E^{jet}_T$ interval. 
The correction factors $F^{MC}_{cal}(r)$ are taken from this tuned version
of PYTHIA. For the comparisons to the measurements presented in Sections~6 and 
7 the untuned version of PYTHIA has been used.

 The correction factors do not show a strong dependence on $\eta^{jet}$ or 
$E^{jet}_T$ and differ from unity by less than 10\% (5\%) for $r \geq 0.4$ 
($r \geq 0.6$). For $r=0.3$ ($r=0.2$) they can increase to 15\% (30\%).
The dependence of the correction factors on the choice of
fragmentation scheme, relative contributions of direct and resolved processes,
fractions of gluon and quark jets in each process, and the inclusion of
multiparton interactions in resolved processes was studied and found to be 
small. The resulting differences are taken into account as contributions to the
total systematic uncertainty assigned to the measurements reported in the
next sections.

\subsection{Systematic uncertainties of the measurements}

 A detailed study of the sources contributing to the systematic uncertainties 
of the measurements was carried out. For each source, the largest change in 
the corrected value of $\psi(r)$ (indicated in parentheses) at a fixed value 
of $r$ typically occurs at $r=0.2$ and the magnitude of the induced change 
decreases rapidly as $r$ increases:
\begin{itemize}
 \item The correction functions to the jet shapes were calculated
       using the event samples of the untuned version of PYTHIA ($\pm$~3\%). 
 \item Using the HERWIG generator to evaluate the energy corrections to 
       the jets and the correction functions to the jet shapes ($\pm$~5\%). 
 \item The correction functions were calculated using Monte Carlo samples of 
       PYTHIA events consisting exclusively of either gluon or quark jets
       ($\pm$~9\%). 
 \item The use of the PYTHIA generator including multiparton interactions in
       resolved processes ($\pm$~4\%). 
 \item Variation of the absolute energy scale of the CAL in the 
       simulated events by $\pm$~3\% ($\pm$~2\%).
 \item The uncertainty in the simulation of the CAL response to low-energy 
       particles ($-$2\%).
 \item Uncertainties in the simulation of the trigger and variation of the
       cuts used to select the data within the ranges allowed by the comparison
       between data and Monte Carlo simulations resulted in negligible
       changes in the corrected jet shapes.
\end{itemize}
The systematic uncertainties for different values of $r$ are correlated. For 
all the results the statistical errors are negligible compared to the 
systematic uncertainties. The total positive (negative) systematic uncertainty
on $\psi(r)$ at each value of $r$ was determined by adding in quadrature
the positive (negative) deviations from the central value. The systematic 
uncertainties were then added in quadrature to the statistical errors and are 
shown as error bars in the figures. 

\section{Inclusive jet photoproduction results}

\subsection{Jet pseudorapidity dependence of the jet shape}

\begin{figure}
\epsfysize=18cm
\epsffile{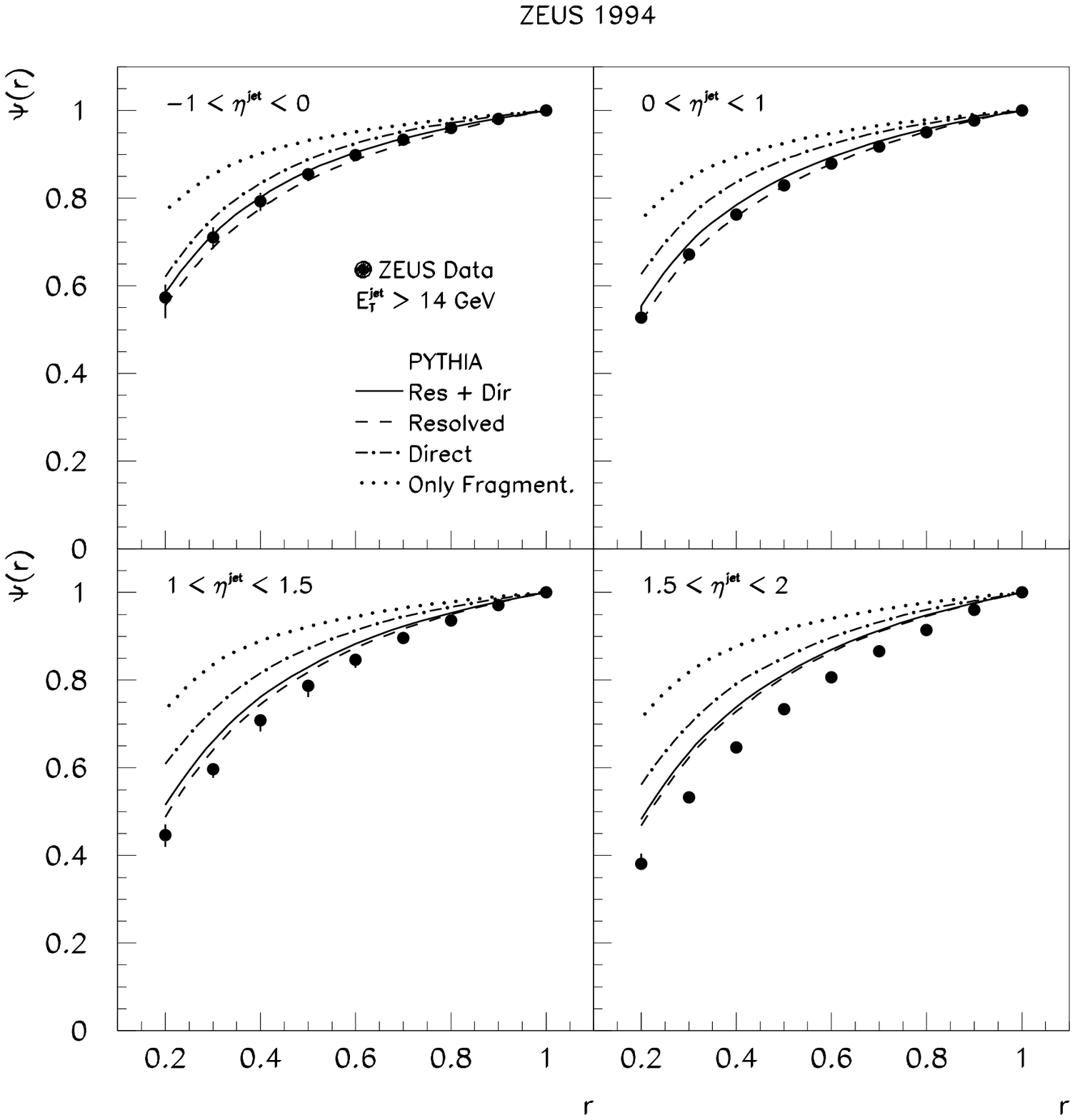}
\caption{\label{figdra1}{ The measured jet shapes corrected to the hadron
 level, $\psi(r)$, for jets in the $E^{jet}_T$ range above 14~GeV 
 in different $\eta^{jet}$ regions.
 The error bars show the statistical and systematic 
 errors added in quadrature. For comparison, the predictions of PYTHIA for 
 resolved (dashed), direct (dot-dashed line), and resolved plus direct 
 processes (solid line) are shown. The predictions of PYTHIA for resolved plus
 direct processes {\rm without} initial and final state parton radiation
 (dotted line) are also included (labelled by `Only Fragment.').}}
\end{figure}

 The jet shape for jets with $E^{jet}_T >$ 14~GeV is presented in 
Figure~\ref{figdra1} for four different $\eta^{jet}$ regions.
It is observed that the jet shape broadens as $\eta^{jet}$ increases. In the
following the predictions of the PYTHIA generator are compared to the
measurements. The predictions using HERWIG are very similar to those of PYTHIA 
and lead to the same conclusions.

\vspace{0.5cm}
\noindent {\bf Comparison to parton shower model predictions}

 The predictions of PYTHIA for direct, resolved, and direct plus resolved 
processes are compared to the measured jet shapes in Figure~\ref{figdra1}. 
The admixture of the two processes was chosen according to the 
cross sections as given by PYTHIA. The dependence of the predictions on the 
specific sets of proton or photon parton distributions is negligible. 
The measured jet shapes in the range $-1<\eta^{jet}<1$ are reasonably well 
described by the predictions of PYTHIA for either resolved plus direct or 
resolved processes alone. The predicted jet shapes in direct processes are 
significantly narrower than those of the data. These results are in agreement 
with the dominance of resolved processes in the $E^{jet}_T$ range studied, 
as observed in the measurement of the inclusive jet 
differential cross sections \cite{zeoct94,h193}. As $\eta^{jet}$ increases 
beyond $\eta^{jet}=1$, the predicted jet shapes increasingly deviate from 
those in the data, which are significantly broader in the most forward region 
($1.5<\eta^{jet}<2$).

 In order to study the relative importance of parton radiation and
fragmentation effects, the predictions of PYTHIA for
resolved plus direct processes {\rm without} initial and final state parton
radiation are also shown in Figure~\ref{figdra1}. Since the predicted jet 
shapes based on fragmentation only are significantly narrower than the 
measured ones, it is concluded that, at the transverse energies studied here, 
the shape of jets is mainly dictated by parton radiation and cannot
be explained by hadronisation alone.

\vspace{0.5cm}
\noindent {\bf Comparison to the parton shower model predictions of 
quark and gluon jets}

\begin{figure}
\epsfysize=18cm
\epsffile{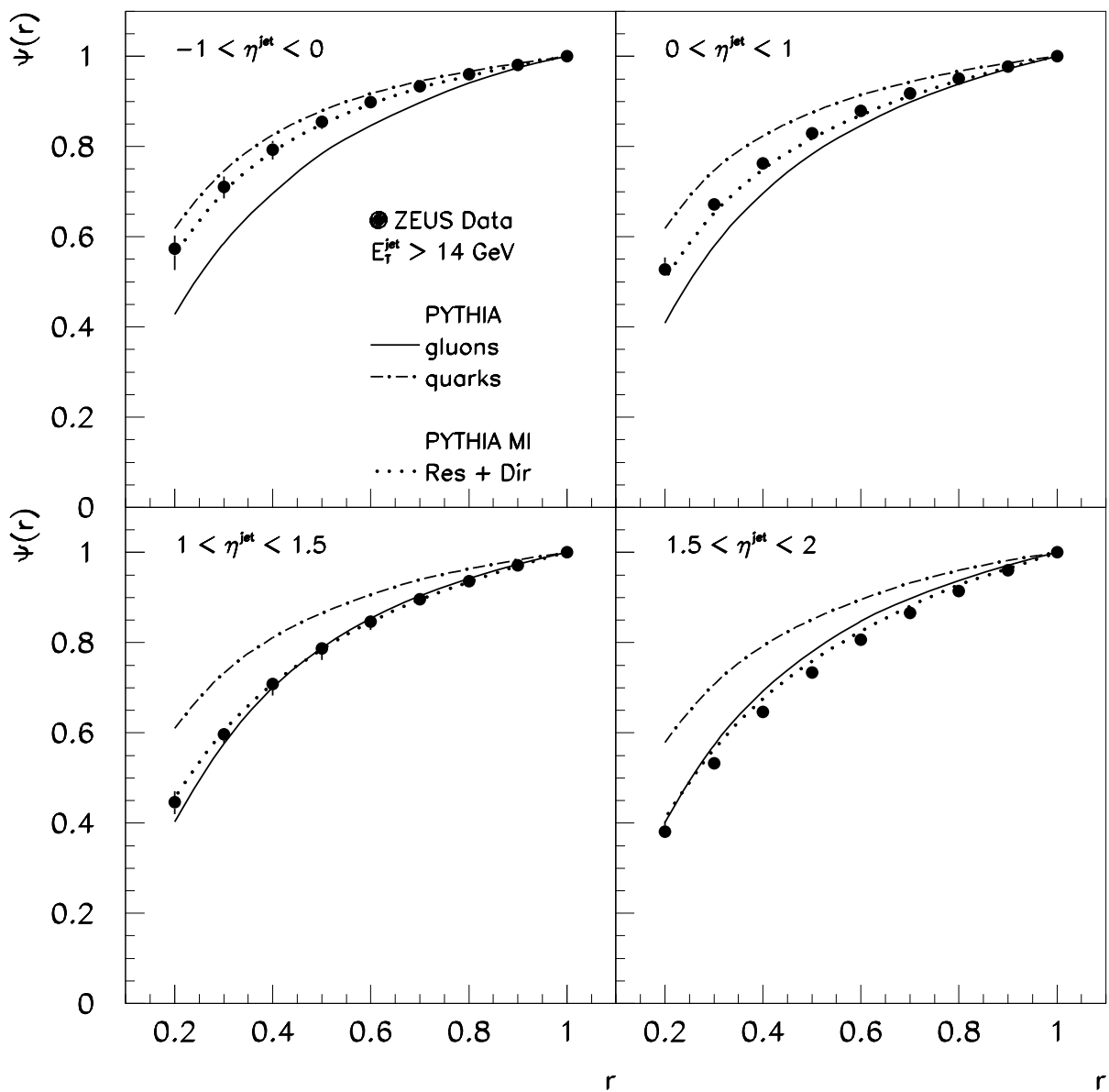}
\caption{\label{figdra2}{ The measured jet shapes corrected to the hadron 
 level, $\psi(r)$, for jets in the $E^{jet}_T$ range above 14~GeV in different
 $\eta^{jet}$ regions. The error bars show the statistical and systematic 
 errors added in quadrature. For comparison, the predictions of PYTHIA 
 including resolved plus direct processes for quark (dot-dashed line) and 
 gluon jets (solid line), and those of PYTHIA MI for resolved plus direct 
 processes (dotted line) are shown.}}
\end{figure}

 The difference between quark and gluon jets is modelled in PYTHIA by a 
leading-logarithm parton-shower approximation. The jet shapes, as predicted by 
PYTHIA, for quark and gluon jets are shown separately in 
Figure~\ref{figdra2}, 
together with the same data as in Figure~\ref{figdra1}. The predictions of 
PYTHIA for the two types of jets are computed separately for resolved and 
direct processes, and then averaged according to the cross sections given by 
PYTHIA. The predicted jet shapes are broader for gluon jets than for quark 
jets in each region of $\eta^{jet}$ and exhibit only a small dependence 
on $\eta^{jet}$.

 The differences in the predictions for resolved and direct processes (see 
Figure~\ref{figdra1}) come mainly from the different fractions of quark and 
gluon jets in the final state depending on the $\eta^{jet}$ region. This 
difference, in turn, originates from the different dominant two-body 
subprocesses. In the case of direct processes, where the dominant subprocess 
is photon-gluon fusion $\gamma g \rightarrow q \bar{q}$, PYTHIA predicts the 
fraction of quark jets to be 80-100\% depending on $\eta^{jet}$. In the case 
of resolved processes, the dominant subprocess in the kinematic regime studied
is $q_{\gamma} g_p \rightarrow q g$ and the predicted fraction of quark
jets has a stronger $\eta^{jet}$-dependence:
from $\sim 80$\% at $\eta^{jet}=-1$ to $\sim 40$\% at $\eta^{jet}=2$. 
The comparison of measurement and prediction (see Figure~\ref{figdra2}) shows 
that the broadening of the jet shapes in the data as $\eta^{jet}$ increases 
is consistent with an increasing fraction of gluon jets. Therefore, one 
possible reason for the deviation between data and prediction is
an underestimation of the fraction of gluon jets in the region $\eta^{jet}>1$ 
by PYTHIA. 

\vspace{0.5cm}
\noindent {\bf Comparison to model predictions including multiparton 
 interactions}

 An excess of transverse energy outside of the jet cone for jets with 
$\eta^{jet}>1$ with respect to the expectations of PYTHIA was observed in 
previous studies \cite{zeoct94,h193}. Since this excess may have some effect 
on the comparisons discussed so far, its influence on the jet shape has been 
considered. In order to simulate an increased energy flow the PYTHIA generator 
including MI was used; the cutoff ($1.0$~GeV) for the effective minimum 
transverse momentum for MI was tuned to reproduce the transverse energy flow 
outside of the jet cone in the data. The jet shapes predicted by PYTHIA MI are
also shown in Figure~\ref{figdra2}. Comparing with Figure~\ref{figdra1},
it is observed that the effects of MI on 
the jet shape are very small in the region $-1<\eta^{jet}<1$, increase 
gradually with $\eta^{jet}$ and yield an improved description of the data in 
the region $\eta^{jet}>1$. The differences in the predicted jet shapes between 
PYTHIA and PYTHIA MI are considered as an estimate of the uncertainty in the 
simulation of the energy flow outside the jet cone due to a possible 
underlying event. The conclusions that the measured jet shapes are consistent 
with the dominance of resolved processes and that the parton radiation is the 
main mechanism responsible for the jet shape still hold when this uncertainty 
is taken into account.

\vspace{1.0cm}
\noindent {\bf Comparison to model predictions for fixed $r$}

\begin{figure}
\epsfysize=18cm
\epsffile{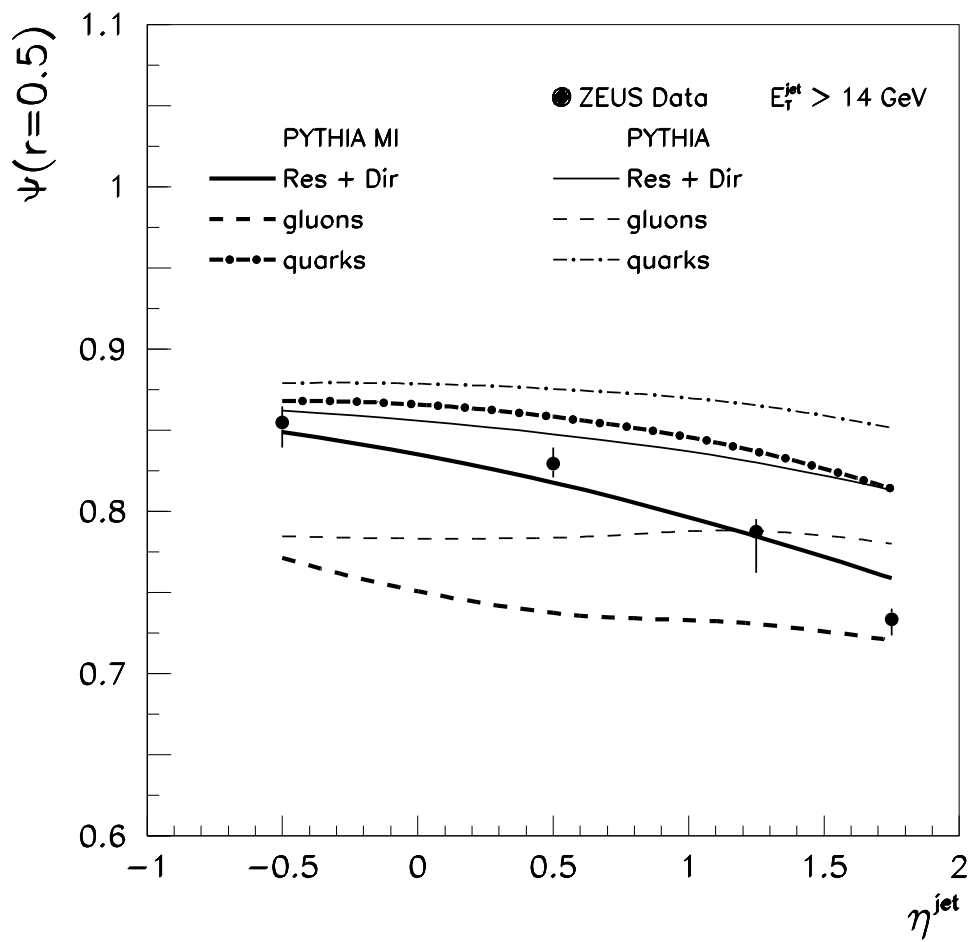}
\caption{\label{figdra3}{ The measured jet shape corrected to the hadron
 level at a fixed value of $r=0.5$, $\psi(r=0.5)$, as a function of 
 $\eta^{jet}$ for jets with $E^{jet}_T$ larger than 14~GeV. The error bars show
 the statistical and systematic errors added in quadrature. For comparison, 
 various predictions of PYTHIA including resolved plus direct processes
 are shown: quark jets (thin dot-dashed line), gluon jets (thin dashed line) 
 and all jets (thin solid line). The corresponding predictions of PYTHIA MI 
 are displayed with thick lines.}}
\end{figure}

Figure~\ref{figdra3} shows the measured jet shape at a fixed value of $r=0.5$,
$\psi(r=0.5)$, as a function of $\eta^{jet}$. The predictions of 
PYTHIA for quark jets with and without MI span a band above the data, while 
those for gluon jets with and without MI span a band typically below the data. 
It is also seen that the effect of MI is larger as $\eta^{jet}$ increases.
The predictions for resolved plus direct processes are also shown in 
Figure~\ref{figdra3}. The prediction of PYTHIA fails to describe the 
relatively strong broadening of the measured jet shape for $\eta^{jet}>1$. As 
mentioned earlier, one reason might be that the fraction of gluon jets in the 
region $\eta^{jet}>1$ is underestimated. However, when the effects of a 
possible underlying event are taken into account, using PYTHIA MI, the 
measured broadening in the forward region is accounted for with the default 
fraction of gluon jets. Note that, in any case, as $\eta^{jet}$ increases the 
measured jet shape changes from a value close to the upper band (quark jets) 
to a value within the lower band (gluon jets). It is concluded that the 
broadening of the measured jet shape as $\eta^{jet}$ increases is consistent 
with an increase of the fraction of gluon jets independent of the effects
of a possible underlying event.

\subsection{$E_T^{jet}$ dependence of the jet shape} 

\begin{figure}
\epsfysize=18cm
\epsffile{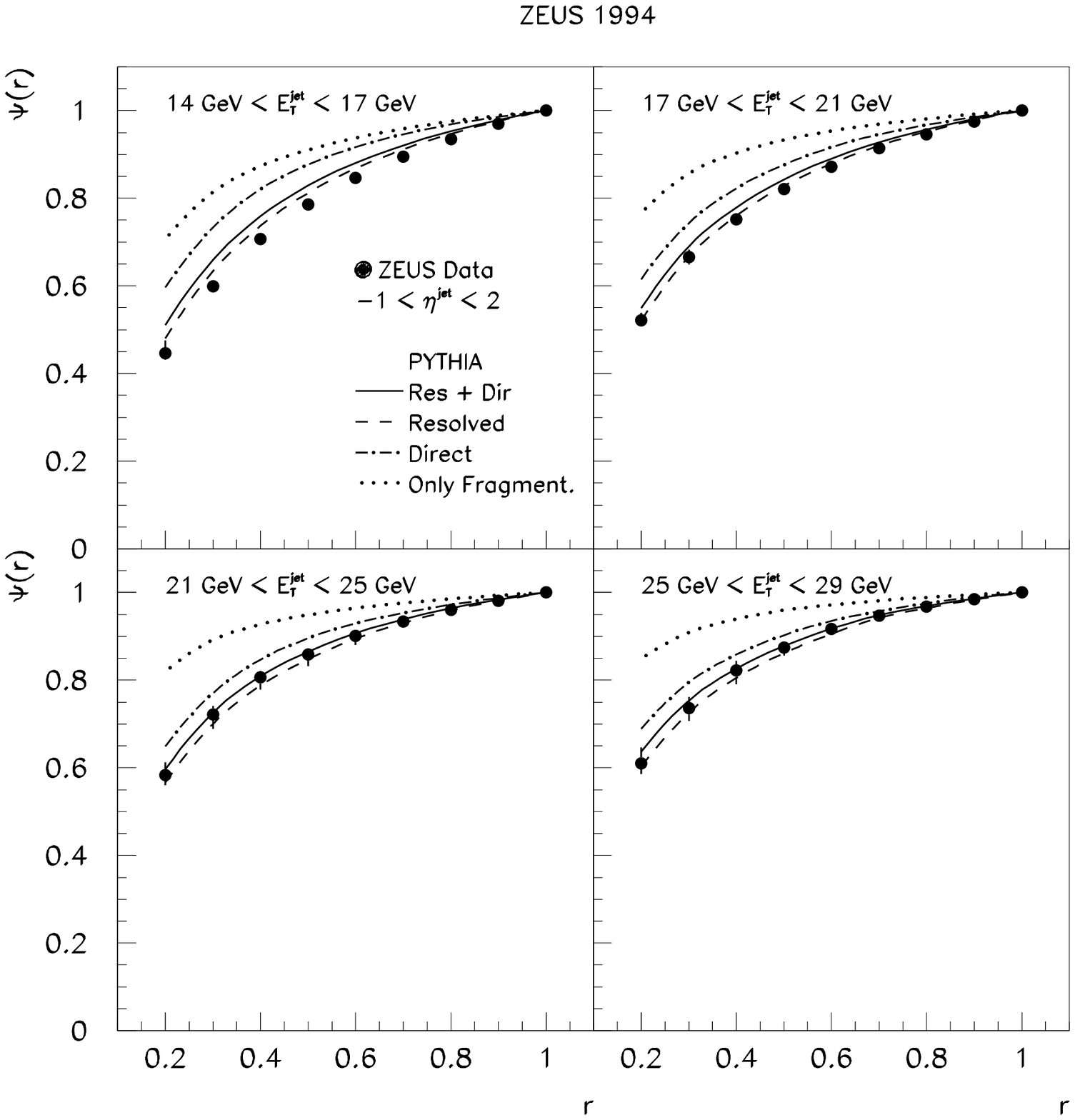}
\caption{\label{figdra4}{ The measured jet shapes corrected to the hadron 
 level, $\psi(r)$, for jets in the $\eta^{jet}$ range between $-1$ and 2 
 in different $E^{jet}_T$ regions. The error bars show the statistical and 
 systematic errors added in quadrature. For comparison, the predictions of 
 PYTHIA for resolved (dashed), direct (dot-dashed line), and resolved plus 
 direct processes (solid line) are shown. The predictions of PYTHIA for 
 resolved plus direct processes without initial and final state parton 
 radiation (dotted line) are also included (labelled by `Only Fragment.').}}
\end{figure}

 The $E^{jet}_T$ dependence of the jet shape is presented in 
Figure~\ref{figdra4}. It is observed that the jets become narrower as 
$E^{jet}_T$ increases. For $E^{jet}_T> 17$~GeV the predictions
of PYTHIA for resolved or resolved plus direct processes reproduce the data 
reasonably well. In the lowest $E^{jet}_T$ region, differences
between data and the predictions are observed. Again the predicted jet shapes 
for direct processes are narrower than for the data. PYTHIA including resolved 
plus direct processes, but {\rm without} initial and final state parton 
radiation, predicts jet shapes which are too narrow in each 
region of $E^{jet}_T$. These comparisons show again that parton radiation is 
the dominant mechanism responsible for the jet shape in the whole range of 
$E^{jet}_T$ studied.

\begin{figure}
\epsfysize=18cm
\epsffile{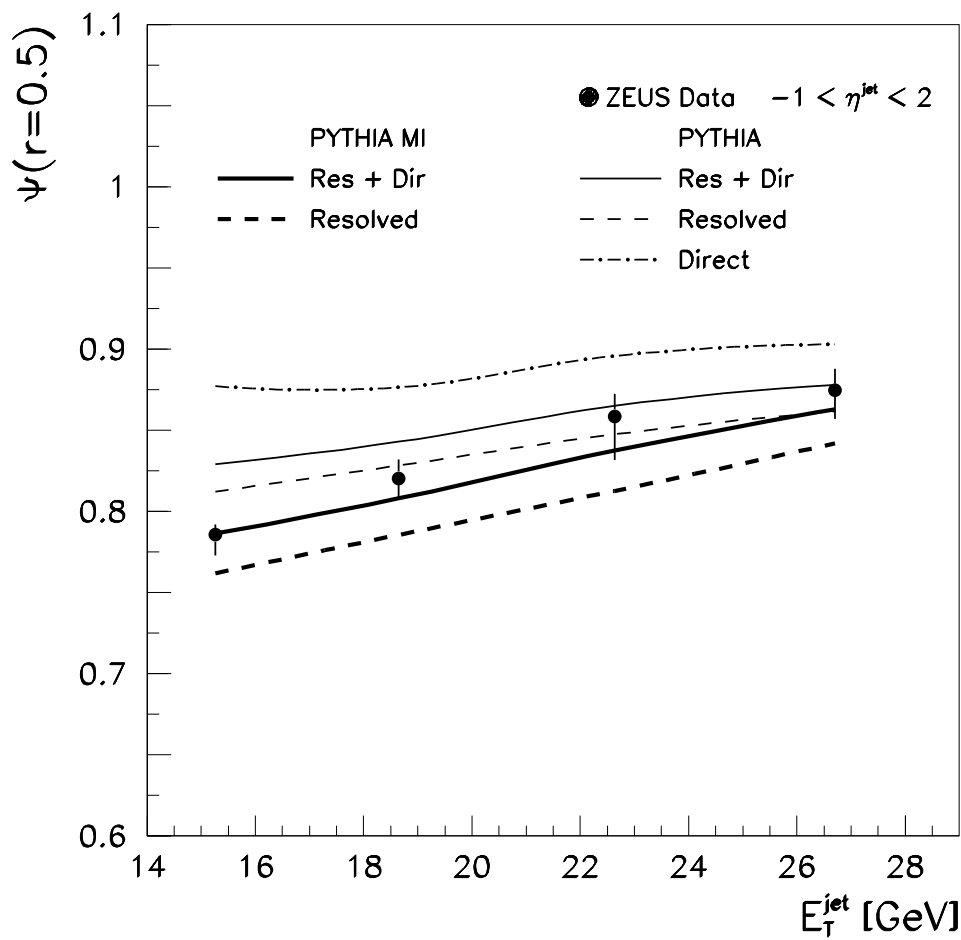}
\caption{\label{figdra5}{ The measured jet shape corrected to the hadron
 level at a fixed value of $r=0.5$, $\psi(r=0.5)$, as a function of $E^{jet}_T$
 for jets in the $\eta^{jet}$ range between $-1$ and 2. The error bars show 
 the statistical and systematic errors added in quadrature. 
 For comparison, various predictions of PYTHIA are shown:
 resolved (thin dashed line), direct (thin dot-dashed line) and resolved plus
 direct processes (thin solid line). The predictions of PYTHIA MI for
 resolved and resolved plus direct processes are
 displayed with thick lines.}}
\end{figure}

 The measured jet shape at a fixed value of $r=0.5$, $\psi(r=0.5)$, shows a 
moderate increase with $E^{jet}_T$, as seen in Figure~\ref{figdra5}. Note that
the jet shapes have been measured in ranges of $E^{jet}_T$ and the data 
points in Figure~\ref{figdra5} are located at the weighted mean in each 
$E^{jet}_T$ range. The predictions for the dependence of the jet shape on 
$E^{jet}_T$ in resolved processes reproduce the data except for the 
lowest $E^{jet}_T$ data point. The predicted jet shape for direct processes 
is narrower than the data. The effects of a possible underlying event are 
estimated using the predictions of PYTHIA MI. 
The inclusion of these effects improves the description of the data in the 
lowest $E^{jet}_T$ data point, but otherwise does not alter 
significantly the $E^{jet}_T$-dependence of the jet shape.

\subsection{Comparison to fixed-order QCD calculations} 

\begin{figure}
\epsfysize=18cm
\epsffile{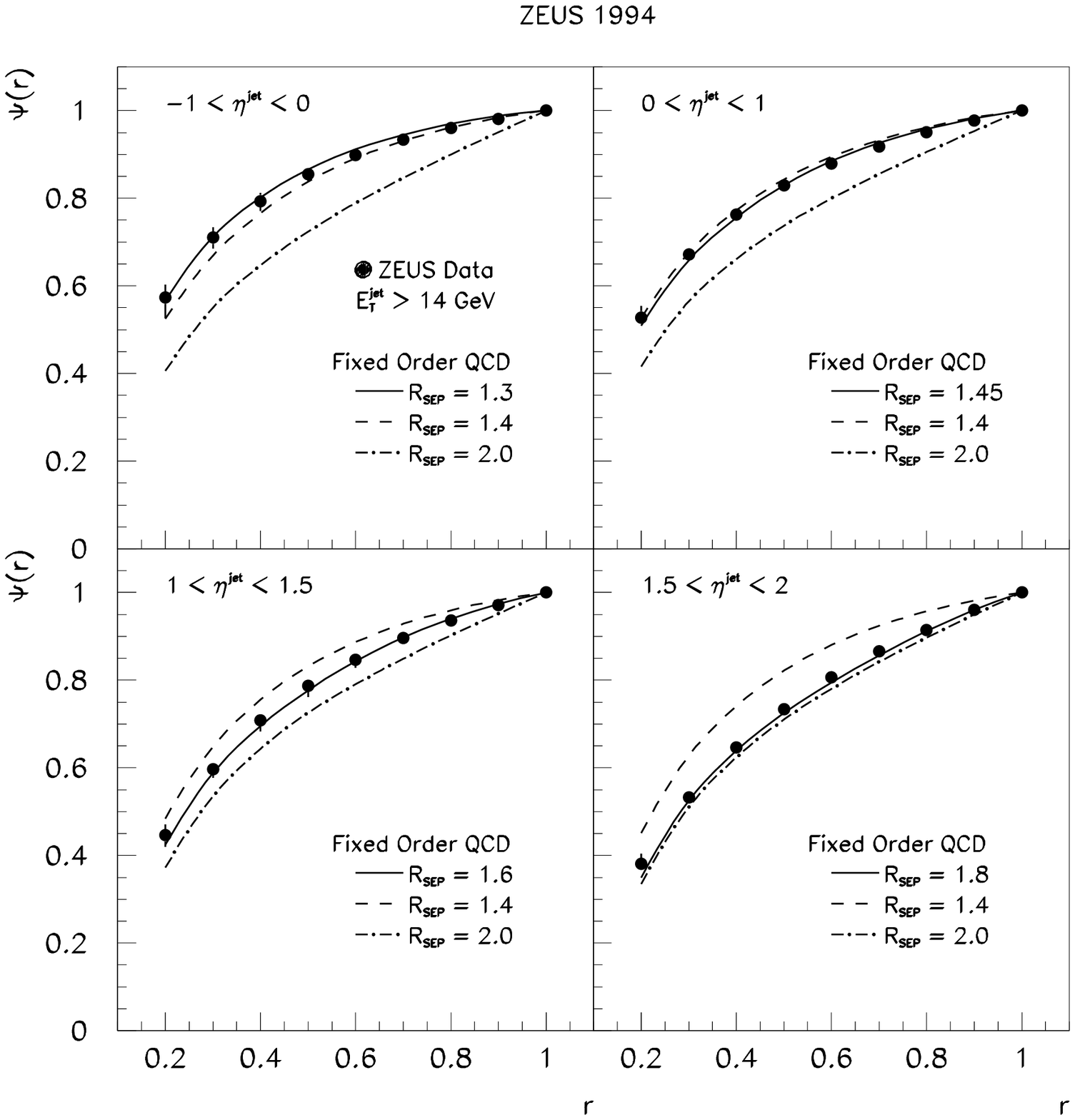}
\caption{\label{figdra6}{ The measured jet shapes corrected to the hadron 
 level, $\psi(r)$, for jets in the $E^{jet}_T$ range above 14~GeV 
 in different $\eta^{jet}$ regions. The error 
 bars show the statistical and systematic errors added in quadrature. For 
 comparison, the predictions for the jet shapes (solid, dashed and
 dot-dashed lines) based upon the fixed-order QCD calculations by M.~Klasen 
 and G.~Kramer with various values of $R_{SEP}$ (see text) are shown.}}
\end{figure}

\begin{figure}
\epsfysize=18cm
\epsffile{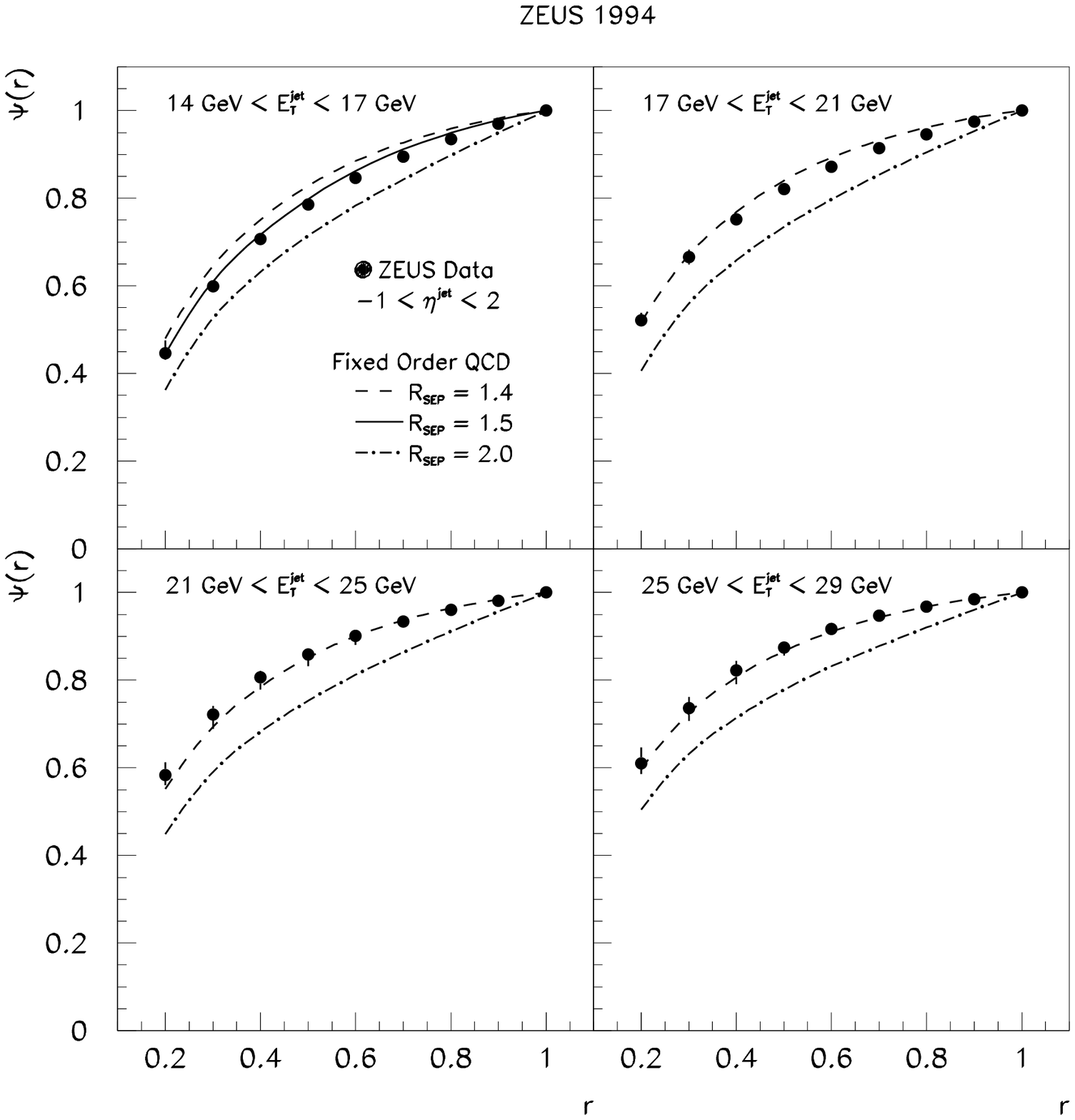}
\caption{\label{figdra7}{ The measured jet shapes corrected to the hadron 
 level, $\psi(r)$, for jets in the $\eta^{jet}$ range between $-1$ and 2 
 in different $E^{jet}_T$ regions. The error bars show the statistical and 
 systematic errors added in quadrature. For comparison, the predictions for 
 the jet shapes (solid, dashed and dot-dashed lines) based upon the 
 fixed-order QCD calculations by M.~Klasen and G.~Kramer with various values 
 of $R_{SEP}$ (see text) are shown.}}
\end{figure}

 Lowest-non-trivial order QCD calculations of the jet shapes 
\cite{kramer,kramerf} are compared to our measurements in 
Figures~\ref{figdra6} and \ref{figdra7}. These predictions include resolved 
and direct processes, and use the CTEQ4 \cite{cteq4} (GRV-HO) proton (photon) 
parton densities. Since the jet shape is computed only to the 
lowest-non-trivial order\footnote{The lowest-non-trivial order contribution
to the jet shape
corresponds to the NLO matrix elements of the hard interaction. 
The NLO contribution to the jet shape
is not available due to the lack of the relevant next-to-next-to-leading order 
matrix elements, and in any case, as pointed out in \cite{seym}, it cannot be
safely calculated for the iterative cone algorithm used here.}, $O(\alpha_s)$,
the predictions are subject to relatively large uncertainties due to the 
strong dependence on the renormalisation and factorisation scales. In the 
calculations shown here, these scales have been chosen equal to $E^{jet}_T$. 
Since the calculations include only up to three partons in the final state, 
not more than two partons can build up a jet. As a result, the overlapping and 
merging issues of the experimental jet algorithm are not reproduced in the 
theoretical calculation \cite{sdellis,giele}. An attempt was made to simulate 
these effects by introducing an ad-hoc $R_{SEP}$ parameter \cite{sdellis}: two
partons are not merged into a single jet if their separation in the 
$\eta-\phi$ plane is more than $R_{SEP}$. The calculations of the jet shape 
shown in Figures~\ref{figdra6} and \ref{figdra7} have been made for various 
$R_{SEP}$ values\footnote{Although the value $R_{SEP}=1$ was suggested 
\cite{workshop} for the comparison between the measurements of cross sections 
and theoretical calculations, other values can be used in order to match the 
measured jet shapes.}: for two fixed values of $R_{SEP}=1.4$ and 
$2.0$, and for the value of $R_{SEP}$ which best reproduces the data 
\cite{kramerf}.

 The fixed-order QCD calculations with a common value of $R_{SEP}=1.4$ 
reproduce reasonably well the measured jet shapes in the region 
$-1<\eta^{jet}<1$ and in the region $E^{jet}_T>17$~GeV. In the forward region, 
$\eta^{jet}>1$, and in the lowest $E^{jet}_T$ region, $14<E^{jet}_T<17$~GeV, 
the calculations with $R_{SEP}=1.4$ show significant deviations with respect
to the data. The discrepancy is very similar to that observed between the 
predictions of PYTHIA without MI and the data. However, a satisfactory 
description of the data can be obtained by leaving $R_{SEP}$ as a free 
parameter (for each interval in $\eta^{jet}$). A crosscheck with the data 
which overlaid jets from different events showed that the minimum distance at 
which two jets are resolved as two distinct jets depends very little on 
$\eta^{jet}$. Therefore, the variation of $R_{SEP}$ with $\eta^{jet}$ cannot 
be justified on these grounds. The use of different values of $R_{SEP}$ to
describe the data in the forward and lowest $E^{jet}_T$ regions may be 
mimicking the effects of QCD higher orders and of a possible underlying event, 
which at present are not included in the calculations. In addition, the 
comparison between the measured jet shapes, which are corrected to the hadron 
level, and the fixed-order QCD calculations at the parton level is
subject to the uncertainty of hadronisation effects.

\section{Dijet photoproduction results}

\begin{figure}
\epsfysize=18cm
\epsffile{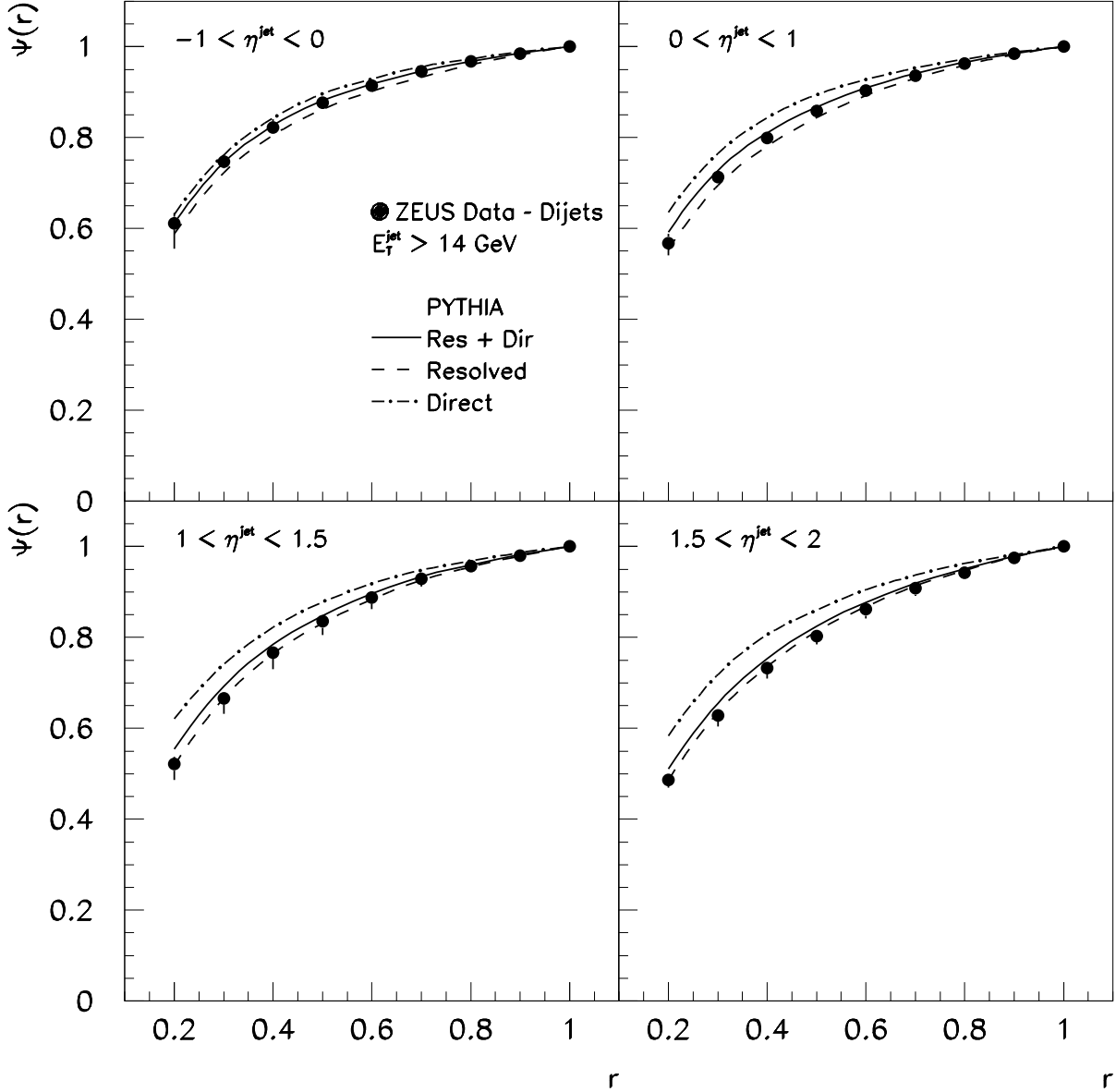}
\caption{\label{figdra8}{ The measured jet shapes corrected to the hadron 
 level, $\psi(r)$, for each of the two highest $E^{jet}_T$ jets in dijet 
 photoproduction. Both jets are required to have $E^{jet}_T>14$~GeV.
 The measurements are shown for four different regions in $\eta^{jet}$. 
 The error bars represent the statistical and systematic errors 
 added in quadrature. For comparison, the predictions of PYTHIA for resolved
 (dashed), direct (dot-dashed line), and resolved plus direct processes 
 (solid line) are shown.}}
\end{figure}

The jet shapes have been measured for each of the two highest $E^{jet}_T$ 
jets in the reaction $e p \rightarrow {\rm jet} \; + \; {\rm jet} \; + \; X$.
Both jets are required to have $E^{jet}_T>14$~GeV and $-1<\eta^{jet}<2$. 
The measurements have been integrated over four non-overlapping regions in 
 $\eta^{jet}$ of each jet. The results for $\psi(r)$ are presented in 
Figure~\ref{figdra8} and are compared to the predictions of PYTHIA. The jet 
shape broadens as $\eta^{jet}$ increases and is narrower than that of the 
inclusive jet sample (see Figure~\ref{figdra1}). 
The predictions for resolved or resolved plus direct processes describe 
reasonably well the measured jet shapes in all regions of $\eta^{jet}$. The 
predicted jet shapes for direct processes are narrower than for the data in 
the region $\eta^{jet}>0$, as expected from the dominance of resolved 
processes in that region. The comparison of the predicted jet shapes with the 
data in the region $\eta^{jet}>1$ does not show the discrepancies 
observed in the inclusive jet photoproduction study. This difference is
attributed to a suppression of the effects of a possible underlying event in 
the case of dijet events since the requirement of two high-$E^{jet}_T$ jets 
increases the contribution from direct processes and, for resolved processes, 
decreases the leftover energy for the remnants. The 
broadening of the measured jet shapes in dijet photoproduction as $\eta^{jet}$ 
increases is adequately reproduced by the predictions of PYTHIA.

 In dijet photoproduction the contributions of resolved and direct processes
can be separated \cite{zenov93,zedij95} by using the variable:
\begin{equation}
 x^{OBS}_{\gamma} = \frac{\sum_{jets} E^{jet}_T e^{-\eta^{jet}}}{2 y E_e},
\end{equation}
where the sum runs over the two jets of highest $E^{jet}_T$ and $y E_e$ is the 
initial photon energy. This variable represents the fraction of the photon's 
momentum participating in the production of the two highest $E^{jet}_T$ jets. 
The LO direct and resolved processes largely populate different regions of 
$x^{OBS}_{\gamma}$, with the direct processes concentrated at high values. 

\begin{figure}
\epsfysize=18cm
\epsffile{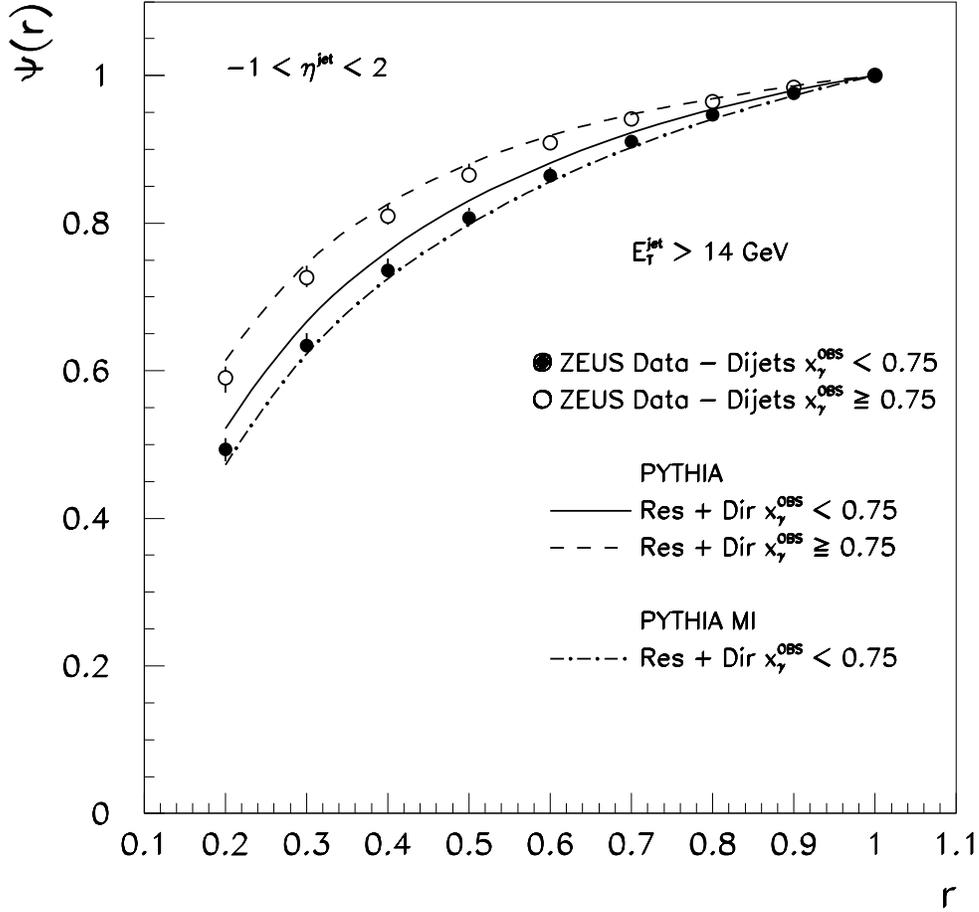}
\caption{\label{figdra9}{ The measured jet shapes corrected to the hadron 
 level, $\psi(r)$, for each of the two highest $E^{jet}_T$ jets in dijet 
 photoproduction separated according to $x^{OBS}_{\gamma}$. Both jets are 
 required to have $E^{jet}_T>14$~GeV and $-1<\eta^{jet}<2$. The error bars 
 show the statistical and systematic errors added in quadrature. For 
 comparison, various predictions of PYTHIA including resolved plus direct 
 processes are shown: for the region $x^{OBS}_{\gamma}<0.75$ (PYTHIA without 
 MI as the solid line and PYTHIA with MI as the dot-dashed line) and for 
 $x^{OBS}_{\gamma} \geq 0.75$ (PYTHIA without MI as the dashed line).}}
\end{figure}

 The results for $\psi(r)$ are presented in Figure~\ref{figdra9} for both
$x^{OBS}_{\gamma}$ smaller and larger than 0.75. It is observed that the 
measured jet shapes for $x^{OBS}_{\gamma} < 0.75$ are broader than those for 
$x^{OBS}_{\gamma} \geq 0.75$. For the region of $x^{OBS}_{\gamma} \geq 0.75$, 
the jet shapes as predicted by PYTHIA including resolved plus direct processes 
describe well the data. The predictions of PYTHIA for the region 
$x^{OBS}_{\gamma}<0.75$ reproduce the data reasonably well though the latter 
are slightly broader. The inclusion of the effects of a possible underlying 
event as modelled with PYTHIA MI leads to an improved description of the 
data in the region of $x^{OBS}_{\gamma} < 0.75$ and has a negligible 
contribution in the region of $x^{OBS}_{\gamma} \geq 0.75$ (not shown).  

\section{\bf Summary and conclusions}

 Measurements of jet shapes in inclusive jet and dijet photoproduction
in $e^+p$ collisions at $\sqrt{s}=300$~GeV using data collected by ZEUS in 
1994 have been presented. The jet shapes refer to jets at the hadron level 
with a cone radius of one unit in the $\eta-\phi$ plane and are given in the 
kinematic region defined by $Q^2 \leq 4$ GeV$^2$ and $134<W<277$~GeV. Jets
with $E^{jet}_T > 14$~GeV and $-1< \eta^{jet} < 2$ have been considered.
The dependence of the jet shapes on $E^{jet}_T$ and $\eta^{jet}$ 
has been studied: the jet shape broadens as $\eta^{jet}$ increases 
and narrows as $E^{jet}_T$ increases. In dijet photoproduction, the jet shapes 
have been measured separately for two regions of $x^{OBS}_{\gamma}$, the 
fraction of the photon's momentum participating in the production of the two 
jets of highest $E^{jet}_T$, $x^{OBS}_{\gamma} < 0.75$ and $x^{OBS}_{\gamma} 
\geq 0.75$. These subsamples are dominated by resolved and direct processes,
respectively. The jet shapes in the region $x^{OBS}_{\gamma}<0.75$ are 
systematically broader than those in the region $x^{OBS}_{\gamma}\geq 0.75$. 

 Leading-logarithm parton-shower Monte Carlo calculations of resolved
and direct processes have been compared to the measured jet shapes. 
The predictions based on resolved or resolved plus direct processes describe 
reasonably well the measured jet shapes in the region $-1<\eta^{jet}<1$ for
inclusive jet photoproduction and in the full region of $\eta^{jet}$ for
dijet production. The predictions including only direct processes are 
narrower than those measured in the data. The removal of initial and final 
state parton radiation in the Monte Carlo calculations gives rise to jet 
shapes which are too narrow compared to those of the data. The results are in 
agreement with the dominance of resolved processes and indicate that parton 
radiation is the dominant mechanism responsible for the jet shape in 
the $E^{jet}_T$ range studied. 

 Fixed-order QCD calculations cannot reproduce the measured jet shapes over 
the full kinematic range with a single value of the $R_{SEP}$ parameter.

 The observed broadening of the jet shape as $\eta^{jet}$ increases 
is consistent with an increase of the fraction of gluon jets independent of 
the effects of a possible underlying event.

\vspace{0.5cm}
\noindent {\Large\bf Acknowledgements}
\vspace{0.3cm}

 The strong support and encouragement of the DESY Directorate has been 
invaluable. The experiment was made possible by the inventiveness and the 
diligent efforts of the HERA machine group.  The design, construction and 
installation of the ZEUS detector have been made possible by the
ingenuity and dedicated efforts of many people from inside DESY and
from the home institutes who are not listed as authors. Their 
contributions are acknowledged with great appreciation. We would like to thank
M. Klasen, G. Kramer and S.G. Salesch for valuable discussions and for
providing us with their calculations.

%--------- REFERENCES -------------

\newpage
\clearpage
%Figure 1
\parskip 0mm

\newpage
\clearpage
%Figure 2
\parskip 0mm

\newpage
\clearpage
%Figure 3
\parskip 0mm

\newpage
\clearpage
%Figure 4
\parskip 0mm

\newpage
\clearpage
%Figure 5
\parskip 0mm

\newpage
\clearpage
%Figure 6
\parskip 0mm

\newpage
\clearpage
%Figure 7
\parskip 0mm

\newpage
\clearpage
%Figure 8
\parskip 0mm

\newpage
\clearpage
%Figure 9
\parskip 0mm

\end{document}